\documentclass[twocolumn]{revtex4}

\input{epsf}

\def\APP{{\it Acta Phys. Pol.} }

\def\IJMP{{\it Int. J. Mod. Phys.} }

\def\JP{{\it J. Phys.} }

\def\NP{{\it Nucl. Phys.} }
\def\PL{{\it Phys. Lett.} }
\def\PR{{\it Phys. Rev.} }
\def\PRL{{\it Phys. Rev. Lett.} }
\def\PRTS{{\it Physics Reports} }

\def\RPP{{\it Rep. Prog. Phys.} }

\def\SP{{\it Soviet. Phys. JETP} }

\def\ZP{{\it Z. Phys.} }

\def\al{\alpha}
\def\be{\beta}
\def\ga{\gamma}
\def\de{\delta}
\def\ep{\epsilon}

\def\ka{\kappa}
\def\la{\lambda}

\def\ta{\tau}

\def\ph{\phi}

\def\om{\omega}

\def\De{\Delta}

\def\La{\Lambda}

\def\Ph{\Phi}

\def\frac#1#2{{\textstyle{{#1}\over
{#2}}}} 
\def\lsim{\mathrel{\rlap{\lower4pt\hbox{\hskip1pt$\sim$}}
    \raise1pt\hbox{$<$}}}
\def\gsim{\mathrel{\rlap{\lower4pt\hbox{\hskip1pt$\sim$}}
    \raise1pt\hbox{$>$}}}
\def\sqr#1#2{{\vcenter{\vbox{\hrule height.#2pt
         \hbox{\vrule width.#2pt height#1pt
\kern#1pt
         \vrule width.#2pt}
         \hrule height.#2pt}}}}

\def\beq{\begin{equation}}
\def\eeq{\end{equation}}
\def\beqa{\begin{eqnarray}}
\def\eeqa{\end{eqnarray}}

\begin{document}

\title{Constraints on Topological Defect Formation in \par First-order
Superconducting Phase Transitions}

\vskip 0.2cm

\author{J. P\'aramos}

\vskip 0.2cm

\affiliation{Instituto Superior T\'ecnico, Departamento de
F\'{\i}sica, \\ Av. Rovisco Pais 1, 1049-001 Lisboa, Portugal
\vskip 0.2cm E-mail address: x\_jorge@netcabo.pt}

\vskip 0.5cm

\begin{abstract}

In this work we address the impact of a cubic term addition to the
Ginzburg-Landau mean-field potential, and study the consequences
on the description of first order phase transitions in
superconductors. Constraints are obtained from experiment and used
to assess consequences on topological defect creation. No
fundamental changes in either the Kibble-Zurek or
Hindmarsh-Rajantie predictions are found.

 \vskip 0.5cm

\end{abstract}

\maketitle

\setcounter{page}{666}


\section{Introduction}

The following work is based on the research reported in Ref.
\cite{Paramos}. It pursues the objective of empathizing the
analogy between accessible condensed matter systems and the
currently accepted framework for the evolution of the Universe,
notably the symmetry-breaking phase transitions \cite{Kibble,
Linde} it has undergone after the Big-Bang. A key aspect to this
comparison is the creation of topological defects, frustrations of
the unbroken phase within the broken one, arising from the
continuity of the order parameter values. These are generally
categorized according to the homotopy group of the quocient of the
unbroken symmetry groups to the broken one and which enables for
comparison of different physical phenomena. These objects can
appear as magnetic monopoles or point-like defects, cosmic
strings, vortices or flux tubes, magnetic domain walls or
textures. Besides its mere aesthetical value, this analogy can
provide a powerful probe into the early stages of the evolution of
the Universe, since direct, hands-on experimental tests are
unattainable: the existence of more accessible systems that
exhibit a formally similar behavior could provide crucial clues to
many cosmologically relevant issues.

These ``cosmology in the laboratory'' experiments can be found in
various systems, ranging from vortices in superfluid phase
transitions of $^4He$ and $^3He$ (see e.g. Ref. \cite{Bunkov,
Vollhardt}), which exhibit common features with cosmic strings
\cite{Zurek}, to liquid crystals undergoing an isotropic-nematic
phase transition \cite{Chuang,Bowick}. Polymer chains were shown
to also possess analogous thermodynamic and transitional behavior
\cite{Bento}. However, most of these systems lack the existence of
a quantity analogous to the magnetic field, which could be a key
player in the early evolution of the Universe and formation of
structure. For that reason, superconductors are a case of special
interest. These comprise phase transitions involving a local gauge
symmetry-breaking process, during which the photon acquires a
``mass'' and, therefore, a penetration length, giving rise to the
Meissner effect: the expulsion of the magnetic field from a
superconducting material, with formation of shielding
``supercurrents'' on its surface.

This symmetry breaking originates topological defects which are
known as flux tubes or vortices, lines of non-null magnetic field
trapped inside the superconductor. Experiments targeted at
observing defect densities in high-$T_c$ materials \cite{Carmi}
were not in accordance with the density predictions of the
Kibble-Zurek (K-Z) mechanism \cite{Kibble}. This, however, is to
be expected, since the K-Z prediction should be accurate only for
global gauge symmetry breaking, when the geodesic rule for phase
angle summation is valid. A new defect generation mechanism, based
on a local gauge treatment by Hindmarsh and Rajantie (H-R)
\cite{Hindmarsh-Rajantie}, leads to an (additive) prediction.
This, although well below the first Carmi-Polturak experimental
sensitivity, is in reasonable agreement with the second.

As a starting point for this research, we note that the above
experiments were both conducted in type-II materials, which
exhibit a higher critical temperature and are therefore easier to
manipulate, leading the current trend in experimental
superconductivity. These materials display a second order phase
transition, with no release of latent heat. On the other hand,
Type-I materials are metastable, showing different responses to a
magnetic field when in normal-superconductor or
superconductor-normal phase transitions. In this work, we try to
account for this more elaborate behavior and to estimate to which
extent it affects the defect density predictions for type-I
superconductors.

Type-I and type-II superconductors are commonly distinguished
according to their Ginzburg-Landau (G-L) parameter $\ka = \la
/\xi$, the ratio between the magnetic field penetration length
$\la$ and the coherence length $\xi$ of the order parameter. In
the presence of a gauge field $\vec{A}$, these characteristic
length scales are obtained from the free energy density

\beq F(\Ph) = {1 \over 2m_e} \left|i \hbar \vec{\nabla} \Ph - {e
\over c} \vec{A} \Ph \right|^2 + V(\Ph) + {1 \over2} \vec{\mu}
\cdot (\vec{\nabla} \times \vec{A})~~, \label{free} \eeq

\noindent where $\vec{\mu}$ is the sample's magnetic moment, $m_e$
is the electron mass and $\Ph$ is the order parameter. The G-L
potential is usually written as \cite{Zurek}

\beq V(\Ph) = \al \Ph^2 + {\be \over 2} \Ph^4~~, \label{pot} \eeq

\noindent where $\al$ is assumed to be linear with temperature,
$\al=\al'(t-1)$, $t \equiv T/T_c$, $\al'$ and $\be$ are constants,
and $T_c$ is the critical temperature. One obtains

\beq \la = \sqrt{{ m_e c^2 \over 4 \pi e^2}{\be \over |\al|}}~~,
\label{lambda} \eeq

\noindent and

\beq \xi = \hbar / \sqrt{2m_e |\al|}~~. \label{xi} \eeq

\noindent

The coherence length at zero temperature is $\xi_0 = \hbar /
\sqrt{2m_e \al'}$, with $ \ka \sim \sqrt{\beta}$. The transition
is second order if $\ka > 1 / \sqrt{2} $, and $\xi_0$ is typically
less than $\sim 0.04 ~ \mu m $; for $\ka < 1 / \sqrt{2} $, the
transition is first order and $\xi_0$ typically greater than $\sim
0.08 ~ \mu m$. It is commonly accepted that, if there is no
applied magnetic field, one always has a second-order phase
transition, for all values of $\ka$. First order transitions arise
from the external field term in Eq. (\ref{free}) if the sample has
a characteristic dimension $l > \la$. This degeneracy of the phase
transition at $H=0$ is, however, only verifiable to the current
experimental sensitivity, and it can be argued that there is some
yet undetected intrinsic metastability, regardless of the applied
magnetic field.

\begin{figure}

\center

\epsfysize=4cm \epsffile{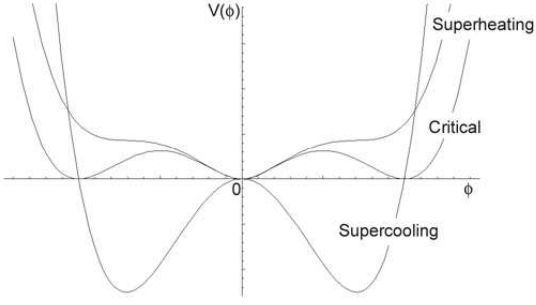}

\caption{Characteristic Potential curves.}

\label{curves}

\end{figure}

Bearing in mind the analogy between condensed matter and
cosmology, we now briefly look at phase transitions in high energy
physics. In thermal field theory (TFT) a first order phase
transition arises due to 1-loop radiative corrections to a
potential similar to that of Eq. (\ref{pot}); a barrier between
minima of the potential is created, as these corrections give rise
to a cubic scalar field term

\beq V(\Ph) = \al \Ph^2 - \ga |\Ph|^3 + {\be\over2} \Ph^4~~,
\label{3pot} \eeq

\noindent where $\ga(T) = (\sqrt{2} / 4 \pi) e^3 T$
\cite{Hindmarsh}. As before, $\be$ is assumed to be constant and
$\al=\al'(t-1)$ to depend linearly with temperature.

In the normal-to-superconductor phase transition, a term similar
to $ - \ga |\Ph|^3 $ also arises if one takes into account gauge
field fluctuations \cite{Halperin, Hove}, producing

\beq \ga = 8 \mu_0 {e \over \hbar c}\sqrt{\pi \mu_0} T_c ~~.
\label{fluct_gamma} \eeq

\noindent This result enables a first order phase transition for
all values of $\ka$. Thermal fluctuations \cite{Hove} and non
local BCS effects \cite{Brandt} describe crossover behavior
between first and second order transitions.

In the following, we adopt a potential of the form of Eq.
(\ref{3pot}) and constrain $ \ga(T) $ based on experimental data.
The results are then compared with both TFT 1-loop radiative
corrections and the the results of Ref. \cite{Halperin} (valid
only at temperature close to $ T_c$, that is, $t \rightarrow 1$).
The introduction of $\ga(T)$ in the potential (\ref{pot}) can
produce changes in both the K-Z and H-R defect generating
mechanisms. Also, a possible nucleation suppression due to the
potential barrier can significantly reduce the number of observed
defects. The obtained constraints on $\ga(T) $ are used to access
the impact on these claims.

\section{Temperature sensitivity bound}

A superconductor undergoing a first-order phase transition crosses
different supercritical fields, displaying a metastable behavior,
as shown in the phase diagram of Figure \ref{curves}. The
superheating curve is given by the condition $ {dV \over d \Ph} =
{d^2 V\over d \Ph^2} = 0 $, for $\Ph \neq 0$, equivalent to $ \al
= 9 \ga^2 / 16 \be $. The supercooling curve is given by the
condition $ {d^2 V(\Ph) \over d \Ph^2} = 0 $ for $\Ph=0$,
corresponding to $ \al = 0 $. The (unobservable) critical curve is
given by $ V(0) = V(\Ph_c) $ and ${dV(\Ph_c) \over d \Ph} = 0$,
where $\Ph_c$ is the non-vanishing  minimum of the potential. This
corresponds to $ \al = \ga^2 /2 \be $.

Assuming $\al = \al' (t-1)$ and $ \ga(t) = \de ~ t$, we obtain for
the superheating curve

\beq \al'(t-1) = {9 \over 16} {\de^2 \over \be}t^2~~,
\label{sh_cond} \eeq

\noindent and

\beq t_{sh} = {2 \over 1 + \sqrt{1 - {9 \over 4} {\de^2 \over \al'
\be}}} \sim 1 + {9 \over 16} { \de^2 \over \al' \be}~~.
\label{t_sh} \eeq

\noindent Due to the presence of the cubic term in the potential
of Eq. (\ref{3pot}), the superheating curve shows a zero-field
shift in temperature from $T_c$ by $ (9 \de^2 / 16 \al' \be) T_c
$. This shift, if detected, would indicate an intrinsic
metastability, in the sense that it does not depend on the
existence of an applied field. Since such temperature shift has
not yet been signaled, the current experimental temperature
sensitivity being $ \De t_{exp} \sim 10^{-3}$ \cite{tempmeas}, a
bound on the slope of $\ga$ is

\beq {9 \over 16} { \de^2 \over \al' \be} < \De t_{exp}~~.
\label{tbound} \eeq

The supercooling transition still occurs at $ t = 1 $, the
critical temperature; this is natural, since it is determined
solely by $\al = 0 $ (neglecting a smaller order correction to
$\al$, \cite{Halperin}).

\section{Superheating perturbation bound}

\begin{table}

\begin{ruledtabular}

\caption{Critical properties of  $Sn$ and $Al$} \label{table1}

\begin{tabular}{|c|c|c|c|c|}

    \hline

Material & $T_c ~(K)$ & $H_c(0)~(G) $ & $ \xi_0 ~(\mu m) $ & $ \la ~(nm) $ \\

\hline \hline

Sn & $ 3.7 $ & $ 309 $ & $ 0.23 $ & $ 34 $ \\

\hline

Al & $ 1.2 $ & $ 105 $ & $ 1.6 $ & $ 16 $ \\

\end{tabular}
\end{ruledtabular}
\end{table}

To obtain further constraints, one looks more carefully at the
mechanism relating the presence of an applied field with
metastability. According to Ref. \cite{Ginzburg}, this arises due
to the contribution of the magnetic moment to the free energy
(\ref{free}), and hence it depends not only on the material (that
is, on $\ka$), but also on its shape and dimensions. The
supercritical fields and the value of $\ka$ have commonly been
obtained from experiments with microspheres of type-I materials.
Table \ref{table1} indicates the critical properties of two of
these, Sn and Al. To assess the influence of the cubic term of Eq.
(\ref{3pot}), we reproduce Ref. \cite{Ginzburg} calculations,
including the cubic term in the potential. For a small sphere of
radius $a$, the magnetic moment is given by \cite{Ginzburg}

\beq {\mu \over V} = -3 \left[1-{3\la \over a \Ph_0}coth{a \Ph_0
\over \la} + {3\la^2 \over a^2 \Ph_0^2}\right] {H\over8\pi}~~,
\label{mu} \eeq

\noindent where $\Ph_0 \equiv \Ph / \Ph_\infty$ and $\Ph_\infty^2
\equiv m_ e c^2 / 4\pi e^2 \la^2$.

After a computation presented in \cite{Paramos}, the reduced
superheating field $h_{sh} \equiv H_{sh} / H_c$ is given by

\beq h_{sh} = \left(1 + {4 \over \sqrt[4]{15}} \ga_G \right)
h_{sh}^0~~, \label{h_sh_gamma}\eeq

\noindent where $\ga_G \equiv 3 \ga / 2 \sqrt{|\al| \be}$ is
defined to be dimensionless, while $h_{sh}^0$ is the unperturbed
superheating field, corresponding to $\ga=0$.

Measurements \cite{data1, data2, data4, data5, data6} were
obtained with colloidal dispersions, and the statistical error due
to the size distributions of the microspheres  do not allow for a
direct fit of $\ga_G$ from $h_{sh}(t)$ data. The measurement
reported in Ref. \cite{data3} used single microspheres, but
non-locality and impurities did lead to a large theoretical
uncertainty in

\beq h_{sh}^0(t) = {1 \over \sqrt{\ka \sqrt{2}}} h_c^0(0) (1 -
t^2)~~, \label{parabola} \eeq

\noindent where $h_c^0(0)=H_c(0)/H_c(t)$, an approximation valid
only close to $T_c$ \cite{hcorr}.

The calculation of the supercooling field implies the evaluation
of a second derivative at the origin, and clearly the presence of
a cubic term in the potential has no effect,

\beq {d^2\over d \Ph^2} (\ga_G \Ph^3) = 6 \ga_G \Ph \rightarrow
0~~. \label{noeffect} \eeq

Since no shift of the superheating curve due to $\ga$  has been
detected, we must have $ \ga_G \ll 1$. This constraint fails for
relative temperatures in the range

\beq 1 - {9 \over 4} {\de^2 \over \al' \be} < t < 1 ~~.
\label{interval} \eeq

\noindent This interval is vanishingly small if

\beq {9 \over 4} { \de^2 \over \al' \be} \ll 1 ~~, \label{hbound}
\eeq

\noindent which is a weaker bound than the one of Eq.
(\ref{tbound}).

Al and Sn show maximum critical fields of order $10^2 ~G$, and the
shift between $h_{sh}$ and $h_{sh}^0$ is smaller than $ 10^{-2} ~G
$. Since this is well below the sensitivity of measurements
\cite{data1, data2, data3, data4, data5, data6}, we consider only
Eq. (\ref{tbound}) and drop the bound of Eq. (\ref{hbound}).
Notice that, even if the ``exclusion" interval'' (\ref{interval})
is non-negligible, no ``spikes" should appear in that region of
the $ H - T $ diagram, due to the smallness of the field values
there.

\begin{table}

\begin{ruledtabular}

\caption{Derived quantities and bounds for $\de$} \label{table2}

\begin{tabular}{|c|c|c|}

Material & Sn & Al\\

\hline

$ \al' ~(J) $ & $ 1.15 \times 10^{-25} $ & $ 2.38 \times 10^{-27} $\\

\hline

$ \be ~(J.m^3) $ & $ 4.72 \times 10^{-54} $ & $ 2.16 \times 10^{-56} $\\

\hline \hline

$ \al' ~(eV^2) $ & $ 3.61 \times 10^{-1}  $ & $ 7.45 \times 10^{-3} $\\

\hline

$ \be $ & $ 9.45 \times 10^{-4} $ & $ 4.32 \times 10^{-6} $\\

\end{tabular}

\vskip 0.5 cm

\begin{tabular}{|c|c|c|c|c|}

bound & $ t_{sh} $ shift & $ \de (eV) $ & $ t_{sh} $ shift & $ \de (eV) $ \\

\hline \hline

$ h_{sh} $ & $ 0.25 $ & $ 1.23 \times 10^{-2} $ & $ 0.25  $ & $ 1.20 \times 10^{-4} $\\

\hline

$ \De T_{exp} $  & $ 10^{-3} $ & $ 7.78 \times 10^{-4} $ & $ 10^{-3}  $ & $ 7.57 \times 10^{-6} $\\

\hline \hline

Ref. \cite{Halperin} & $ 5.19 \times 10^{-9} $ & $ 1.77 \times 10^{-6} $ & $ 2.32 \times 10^{-6}  $ & $ 3.64 \times 10^{-7} $\\

\hline

TFT & $ 2.92 \times 10^{-9} $ & $ 1.33 \times 10^{-6} $ & $ 3.24 \times 10^{-6}  $ & $ 4.31 \times 10^{-7} $\\

\end{tabular}
\end{ruledtabular}
\end{table}

Table \ref{table2} provides a comparison of the bounds on $\de$
with the cubic term arising from 1-loop corrections in TFT and the
prediction of Ref. \cite{Halperin}. This is achieved by computing
the slope of $\ga(t)$ from $\ga(T) = (\sqrt{2} / 4 \pi) e^3 T $,
obtaining $ \de = (\sqrt{2} / 4 \pi) e^3 T_c $. Notice that
$\ga(t)$ as a function of the reduced temperature is material
dependent, although $\ga(T)$ is not. The quantities $\al'$ and
$\be$ are also included.

We notice that unit conversion is not direct, but achieved through
a multiplicative factor $m_e$: since the dimension of the scalar
field in G-L theory is $[\Ph^2]=L^{-3}$, its square representing a
density, while in field theory $[\Ph]=L^{-1}$, the dimensionality
of $\ga$ depends of the theory at hand. For comparison sake, we
have chosen $[\ga] = L^{-1}$. Hence, the definition of $\ga$ is
changed with respect to the free energy potential of Eq.
(\ref{3pot}), through a convenient $m_e$ factor; the electron mass
determines the conversion as it is absent from the kinetic term of
the Lagrangean density of field theory, $
\partial_\mu\Ph \partial^\mu\Ph $, but present in the corresponding condensed
matter free energy term, $ (\hbar^2 / 2 m_e) \nabla^2\Ph$;
equivalently, one can loook at the coherence length: $\xi^2_{FT} =
1/ \al'$ \textit{vs.} $\xi^2_{cm} = \hbar^2 / 2 m_e \al'$.

According to Ref. \cite{Halperin}, in the absence of an applied
magnetic field, momentum fluctuations of the gauge field have an
expectation value derived from the equipartition theorem.
Integrating over momentum space (with a cutoff $\La$ of the order
of $\xi_0^{-1}$), one gets

\beq \langle A^2\rangle _\Ph = 4 {\mu_0 \over \pi} \La T_c - 8
\mu_0 {e \over \hbar c} \sqrt{\pi \mu_0} T_c |\Ph|~~.
\label{meangauge} \eeq

\noindent Hence, from this result arises, asides from an
unimportant correction to the scalar field mass (or coherence
length), a more relevant (negative) cubic term, $- 8 \mu_0 (e /
\hbar c)\sqrt{\pi \mu_0} T_c |\Ph|^3~$. At zero field, this
produces a shift in the superheating temperature of

\beq \De_T = 7.25 \times 10^{-12} T_c^3 H_c(0)^2 \xi_0^6
~~,\label{DeltaT} \eeq

\noindent with $H_c(0)$ expressed in \textit{Gauss} and $\xi_0$ in
$\mu m$. For Sn, it requires a temperature sensitivity of $10^{-9}
K$, well below current possibilities. Al, however, requires just a
sensitivity of $10^{-6} K$, attainable if one employs state of the
art relative temperature measurement techniques.

One can see that, for both materials, the slopes of $\ga$
predicted by TFT and Ref. \cite{Halperin} have similar magnitudes
$\sim 10^{-7} eV$. This is a confirmation of the underlying
analogous mechanisms: one can view the thermal averaging of the
gauge field in condensed matter as equivalent to finite
temperature vacuum polarization, given by renormalization of
1-loop Feynmann diagrams.

\section{Topological defect formation}

We now discuss possible implications of the presence of this cubic
term in the mean-field potential: the K-Z mechanism predicts a
topological defects (vortices) density of $n \simeq
\xi^{-2}_0({\ta_0/\ta_q})^\nu$, where $\ta_0 = \pi \hbar / 16 k_B
T_c $ is the characteristic time scale, given by the Gorkov
equation, $\ta_q$ is the quench time, and $\nu$ is a critical
exponent. One topological defect per $\xi^2_0$ area is assumed.
Since the characteristic scales $\xi_0$ and $\la$ are obtained by
linearizing the G-L equations close to $T_c$, thus neglecting the
$\ga$-cubic and $\be$-quartic terms contribution, we expect no
significant changes to $\xi_0$ and, hence, to this prediction.

The defect density predicted by the H-R mechanism for a thin slab
of width $L_z$ is given by $n \simeq (e/2 \pi) T^{1/2} L_z^{-1/2}
\hat{\xi}^{-1}$, where $\hat{\xi} \sim 2 \pi/\hat{k}$ is the
domain size after the transition and is related to the highest
wave number $\hat{k}$ to fall out of equilibrium. This is obtained
from the adiabaticity relation

\beq \left\vert{d\om(k) \over dt}\right\vert = \om^{2}(k)~~.
\label{dispersion} \eeq

For an underdamped dispersion relation, one has
$\om(k)=\sqrt{k^2+m_\ga^2}~$ and the photon mass is given by
$m_\ga^2 = 2e^2 |\Ph|^2 = -2e^2 \al / \be~$. This leads to a
defect prediction $n\propto\ta_q^{- 1/3}~$. Recall that the
penetration length $\la$ expresses a non-null photon ``mass''
arising from the spontaneous symmetry breaking that occurs during
the normal-to-superconductor phase transition. The presence of a
cubic term in the potential will change the photon mass, as the
true vacuum of the broken phase shifts to

\beq \Ph = {-3\ga+\sqrt{-16\al\be+9\ga^2}\over4\be}~~.
\label{field} \eeq

\noindent However, since $\ga_G \equiv 3 \ga / 2 \sqrt{|\al|\be}
\ll 1 $, the effect is, as in the K-Z case, too small to affect
the H-R prediction in a significant way.

There is also a non-vanishing probability of the order parameter
to quantum tunnel from the false vacuum towards the true one. The
rate of transition per unit volume and time is given, in the thin
wall approximation, \cite{Coleman, Linde2}

\beq {\Gamma \over V \Delta t} = T^4\left({S_3 \over 2
\pi~T}\right)^{3/2}e^{-S_3/T}~~. \label{decayingrate} \eeq

\beq S_3(T) = {2\pi \over 81} {1 \over \be^7 \sqrt{\be}} {\ga^9(T)
\over \ep^2(T) }~~ \label{thinwall} \eeq

\noindent is the Euclidean action, and $\ep(T)$ the ``depth" of
the true vacuum. For the thin wall approximation to be valid, one
assumes that the barrier's height is much greater than this
``depth'' $\ep$, which implies that $\ga$ is comparable to $\al$
and $\be$, that is, $\ga_G \sim 1$. Since, as shown above, the
current temperature sensitivity of $ 10^{-3} K $ only allows for $
\ga_G < 10^{-2}$, the approximation breaks: the field should
always tunnel through the potential barrier, that is, with a
probability close to unity. Therefore, topological defect
production is unsuppressed, and one does not need to be concerned
that these may not have sufficient time to nucleate within the
measurement's resolution time.

\section{Conclusions}

In this work we have examined a description of a type-I
superconductive phase transition including a cubic term in the G-L
mean-field potential, inspired both by analogy with TFT 1-loop
corrections and gauge field thermal averaging \cite{Halperin}.
Superheating field and temperature constraints impose bounds on
this cubic term so that its effects are small compared to that of
other parameters in the G-L potential. In particular, there is
negligible impact on the defect density predictions of K-Z
\cite{Kibble,Zurek} or H-R \cite{Hindmarsh-Rajantie}, with no
suppression or slowing down of defect production due to nucleation
suppression arising from the induced potential barrier between
false and true vacua.

In the absence of any sizeable perturbations, the H-R prediction
for topological defect density in type-I superconductors should be
reduced by a factor of $10-100$. However, it has been suggested
that defects nucleated in type-I materials survive significantly
longer than in type-II \cite{Shapiro}: for type-I superconductors,
their lifetime is expected to be of order $10^{-4}$ seconds. This
increases the possibility of measuring the created defects before
they disappear, possibly compensating for an inferior net number.

Finally, we point out that the superheating temperature shift
induced by a cubic term, derived either from Ref. \cite{Halperin}
or from TFT 1-loop corrections, increases with decreasing $\ka$
($\De t_{sh}(Sn) \sim 10^{-9}$; $\De t_{sh}(Al) \sim 10^{-6}$).
Thus, possible searches for a cubic term should be conducted with
extreme ($\ka \ll 1$) type-I materials like $\al$-tungsten, which
displays $ T_c = 15.4 \pm 0.5 ~mK $, $ H_c = 1.15 \pm 0.03 ~G $.
Also, much more precise bounds could be attained by using a DC
SQUID to measure the shift in the supercritical field, since this
device currently possesses a sensitivity of $10^{-5} \ph_0 /
\sqrt{Hz} $, or $10^{-6}G $ over a 10 $\mu$m grain diameter.

\begin{acknowledgments}

\noindent This talk was presented at the COSLAB Workshop
``Cosmological Phase Transitions and Topological Defects'', which
took place in 22-24 May 2003, at Porto, Portugal. The author
thanks the organisers for their hospitality, and also Tom Girard
and Orfeu Bertolami for remarks and helpful discussion. The work
was supported by grants PRAXIS/10033/98 and CERN/40128/00 of the
Portuguese Funda\c{c}\~{a}o para a Ci\^{e}ncia e Tecnologia. JP is
sponsored by the Funda\c{c}\~{a}o para a Ci\^{e}ncia e Tecnologia
under the grant BD~6207/2001.

\end{acknowledgments}

\end{document}